%% file: SiC-chargecontrol.tex
\documentclass[prl,twocolumn,superscriptaddress,amsmath,amssymb]{revtex4-1}

\input{Defs.tex}

\begin{document}

\title{Optical charge state control of spin defects in 4H-SiC}

\author{Gary Wolfowicz}
\affiliation{Institute for Molecular Engineering, University of Chicago, Chicago, Illinois 60637, USA}
\affiliation{WPI-Advanced Institute for Materials Research (WPI-AIMR), Tohoku University, Japan}

\author{Christopher P. Anderson}
\affiliation{Institute for Molecular Engineering, University of Chicago, Chicago, Illinois 60637, USA}
\affiliation{Department of Physics, University of Chicago, Chicago, Illinois 60637, USA}

\author{Andrew L. Yeats}
\affiliation{Institute for Molecular Engineering, University of Chicago, Chicago, Illinois 60637, USA}
\affiliation{Materials Science Division, Argonne National Laboratory, Lemont, Illinois 60439, USA}

\author{Samuel J. Whiteley}
\affiliation{Institute for Molecular Engineering, University of Chicago, Chicago, Illinois 60637, USA}
\affiliation{Department of Physics, University of Chicago, Chicago, Illinois 60637, USA}

\author{Jens Niklas}
\affiliation{Chemical Sciences and Engineering Division, Argonne National Laboratory, Lemont, Illinois 60439, USA}

\author{Oleg G. Poluektov}
\affiliation{Chemical Sciences and Engineering Division, Argonne National Laboratory, Lemont, Illinois 60439, USA}

\author{F. Joseph Heremans}
\affiliation{Materials Science Division, Argonne National Laboratory, Lemont, Illinois 60439, USA}
\affiliation{Institute for Molecular Engineering, University of Chicago, Chicago, Illinois 60637, USA}

\author{David D. Awschalom}
\email{awsch@uchicago.edu}
\affiliation{Institute for Molecular Engineering, University of Chicago, Chicago, Illinois 60637, USA}
\affiliation{Materials Science Division, Argonne National Laboratory, Lemont, Illinois 60439, USA}

\date{\today}

\begin{abstract}
Defects in silicon carbide (SiC) have emerged as a favorable platform for optically-active spin-based quantum technologies. Spin qubits exist in specific charge states of these defects, where the ability
to control these states can provide enhanced spin-dependent readout and long-term charge stability of the qubits. We investigate this charge state control for two major spin qubits in 4H-SiC, the divacancy (VV) and silicon vacancy (\Vsi), obtaining bidirectional optical charge conversion between the bright and dark states of these defects. We measure increased photoluminescence from VV ensembles by up to three orders of magnitude using near-ultraviolet excitation, depending on the substrate, and without degrading the electron spin coherence time. This charge conversion remains stable for hours at cryogenic temperatures, allowing spatial and persistent patterning of the relative charge state populations. We develop a comprehensive model of the defects and optical processes involved, offering a strong basis to improve material design and to develop quantum applications in SiC.
\end{abstract}

\maketitle

Optically active color centers in wide bandgap semiconductors have shown considerable potential for a variety of spin-based quantum technologies, from quantum computing and quantum memories~\cite{Waldherr2014} to nano-scale sensing~\cite{Maze2008a, Toyli2013,Kucsko2013}. Spin defects in silicon carbide (SiC) in particular combine the optical properties required for single-spin measurements~(\cite{Baranov2011, Koehl2011,Kraus2013,Christle2014,Widmann2015,Christle2017}) with wafer-scale growth and silicon-like fabrication capabilities developed for high-power electronics. However, optimizing these systems for spin qubit applications requires an understanding of not only their spin and optical properties, as demonstrated in the negatively charged silicon vacancy (\Vsim)~\cite{Janzen2009,Baranov2011} and the neutral divacancy (\VVO)~\cite{Torpo2002,Son2006,Falk2013a,Christle2017}, but also an understanding of their charge properties.

Impurities in SiC and their charge states have been investigated for conventional electronics applications, as they play an important role in transport properties and in carrier compensation. Most studies involve deep level transient spectroscopy (DLTS)~\cite{Booker2014,Booker2016}, electron spin resonance (ESR)~\cite{Matsumoto1997,Isoya2008} and density functional theory (DFT)~\cite{Gali2012,Gordon2015,Weber2011} with a strong focus on the carbon vacancy (\Vc)~\cite{Umeda2005,Son2012,Booker2016}; fewer works have addressed \Vsi\ and VV defects~\cite{Umeda2009}. For the purpose of quantum information, it is desirable to understand the complete physics of the defects themselves, not just their influence on transport or other electrical characteristics of the substrate.

Here, we investigate the effect of optical illumination on the stability of the relevant (optically bright) charge states of VV and \Vsi, the ability to control and convert these states between different charge levels, and the implications for quantum applications. We investigate these questions using a combination of techniques including photoluminescence (PL), optically-detected magnetic resonance (ODMR) and electron spin resonance (ESR). The VV and \Vsi\ charge states are both stabilized to the \VVO\ and \Vsim\ states required to observe PL, whose intensity can be enhanced by up to three orders of magnitude depending on the material (local defect concentrations and Fermi level). For VV in particular, we observe bidirectional charge conversion between the neutral (bright qubit state) and negative charge states using mainly near-ultraviolet (365-405~nm) and near-infrared (976~nm) light. This charge conversion is stable at cryogenic temperature and does not affect the ODMR contrast nor the electron spin coherence time, and can therefore be readily applied to increase PL emission from ensembles. 

Charge state conversion can have multiple origins, including direct photoionization, free carrier recombination, and charge transfer between defects. In order to fully understand the involved processes, we measure the charge dynamics of VV, \Vsi\ and nitrogen (N) under illumination, where N is the main dopant in our semi-insulating 4H-SiC samples. Excitation dependence with wavelengths ranging from 365~nm to 1310~nm were measured and simulated, offering a comprehensive picture of charge transfer between these defects. In particular, this allows us to identify that \VVO\ converts to the dark \VVm\ charge state under 976~nm illumination, while \Vsim\ will convert to the dark \VsiO\ charge state with above-bandgap light.

Control and understanding of these charge dynamics is crucial for maximizing spin qubit readout, choosing adequate background impurity concentrations in samples and optimizing designs of SiC nano-devices for quantum applications. Such methods have also been applied in the nitrogen-vacancy (NV) center in diamond for quantum optics applications~\cite{Aslam2013}, enabling for example reduced spectral diffusion~\cite{Siyushev2013} or Stark tuning of the optical transitions through photoexcitation of trapped charges \cite{Bassett2011}. More exotic applications of charge dynamics include high density data storage~\cite{Dhomkar2016}, STED super-resolution imaging~\cite{Han2010,Chen2015} and charge quantum buses~\cite{Doherty2016}.
 
\section{Results}
\subsection{PL enhancement using UV illumination}
\begin{figure}[t]
	\centerline{\includegraphics[width=\figwidth]{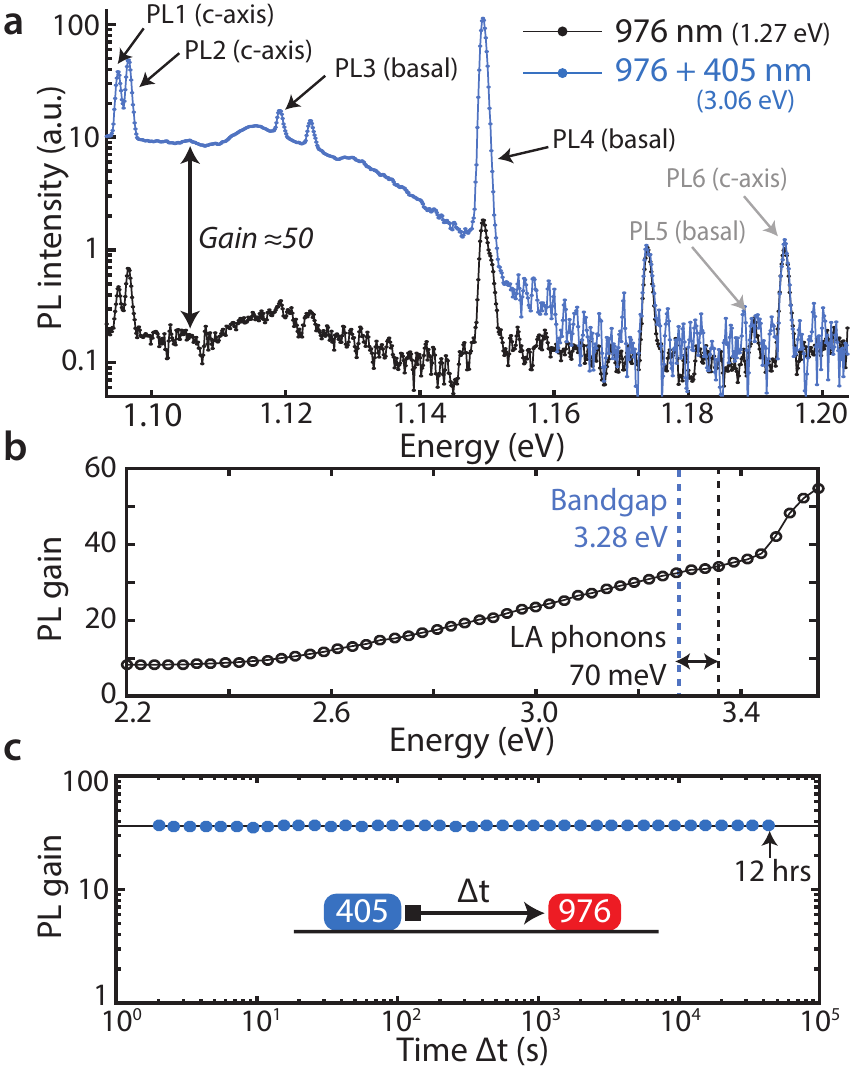}}
	\caption{\textbf{Effect of near-bandgap illumination on 4H-SiC divacancies.} 
		{\bf (a)} PL spectrum with 976~nm excitation of the various divacancies in 4H-SiC, as designated in~\cite{Koehl2011}, without and with continuous illumination at 405~nm ($\approx 5$~mW optical power). All the observed PL lines except PL5 and PL6 are enhanced by the UV excitation, including both c-axis and basal defects.
		{\bf (b)} Gain in the PL signal (integrated across PL1-4) as a function of excitation wavelength (energy) around the 4H-SiC bandgap (3.28~eV at 5~K~\cite{Galeckas2002}). Power was normalized to 0.4~$\mu$W across the entire energy range. The onset of change in the curve is shifted from the bandgap energy due to absorption of longitudinal acoustic phonons (about 70-80~meV)~\cite{Galeckas2002}.
		{\bf (c)} Lifetime of the \VVO\ charge state after a 405~nm pulse at 6~K. No significant decay is observed after 12~hours.
	} 
	\label{fig:PLspectrum}
\end{figure}

We initially observe a drastic increase in PL intensity of \VVO, by about 50 times, when continuously illuminating a semi-insulating 4H-SiC sample with a 405~nm ("UV") laser diode, in addition to the 976~nm laser required for PL excitation. This is shown in \Fig{fig:PLspectrum}(a) where the full PL spectrum for all the divacancies (PL1 to PL6~\cite{Koehl2011}) is taken with (blue) and without (black) 405~nm excitation. Both c-axis (PL1, PL2) and basal defects (PL3, PL4) show an increase in their PL intensity, with slight variation between defects, which we ascribe to charge conversion of the divacancy toward its observable neutral state. On the other hand, PL5 and PL6 remain completely unaffected, adding another unique feature to these currently unidentified defects on top of their strong room temperature PL emission. The \VVO\ PL enhancement with UV was observed in all semi-insulating wafers we measured, with gains ranging by a factor of 2 to 1000 (see Supplementary Figure 2), including samples obtained from separate commercial suppliers (Cree or Norstel), different growth batches, or even simply from separate positions within the same wafer. This strongly indicates an influence from the local environment, e.g. from the remaining concentration of N dopants or other impurities which is known to locally differ in as-grown wafers~\cite{Jenny2004}. The PL intensity with UV remains fairly constant however from sample to sample.

In order to understand the effect of 405~nm illumination, and optimize the enhancement, the excitation wavelength is swept across the 4H-SiC bandgap energy (3.28 eV, 380~nm) as shown in \Fig{fig:PLspectrum}(b). The PL gain slowly increases with excitation energy, and around 3.33-3.35~eV, slightly above the 4H-SiC bandgap (3.28~eV), it drastically turns up. This suggests two separate processes are altering the VV charge state from either \VVm\ or \VVp\ to \VVO: at low energies, we will see this is due to direct photoionization, while at high energies, the gain results from recombination of generated electron-hole pairs.

We now consider charge dynamics under illumination, starting from the stability of the conversion observed after UV excitation. As illustrated in \Fig{fig:PLspectrum}(c), the system is initially pumped with 405~nm toward a high \VVO\ population (strong PL intensity), followed by a long delay to allow for relaxation and finally measurement using 976~nm. No change is observed over the course of 12~hours, a result largely expected for a deep defect at cryogenic temperature (6~K). More interestingly, the PL intensity always drops to a low level after turning off the UV excitation while 976~nm was continuously on. Combined with this long stability, this implies that the use of 976~nm to excite VV PL is simultaneously converting the VV out of its neutral charge state, toward a dark state (more details are given later on). This has significant consequences as wavelengths near 976~nm have been extensively used in recent PL- and ODMR-related works \cite{Christle2014,Zargaleh2016,Seo2016}, owing to being close to the absorption maximum of the ground to excited state transition of \VVO, as well as being easily available commercially. These previous studies may therefore have been partially perturbed by charge conversion.

\subsection{Illumination effects on spin properties}
\begin{figure}[t]
	\centerline{\includegraphics[width=\figwidth]{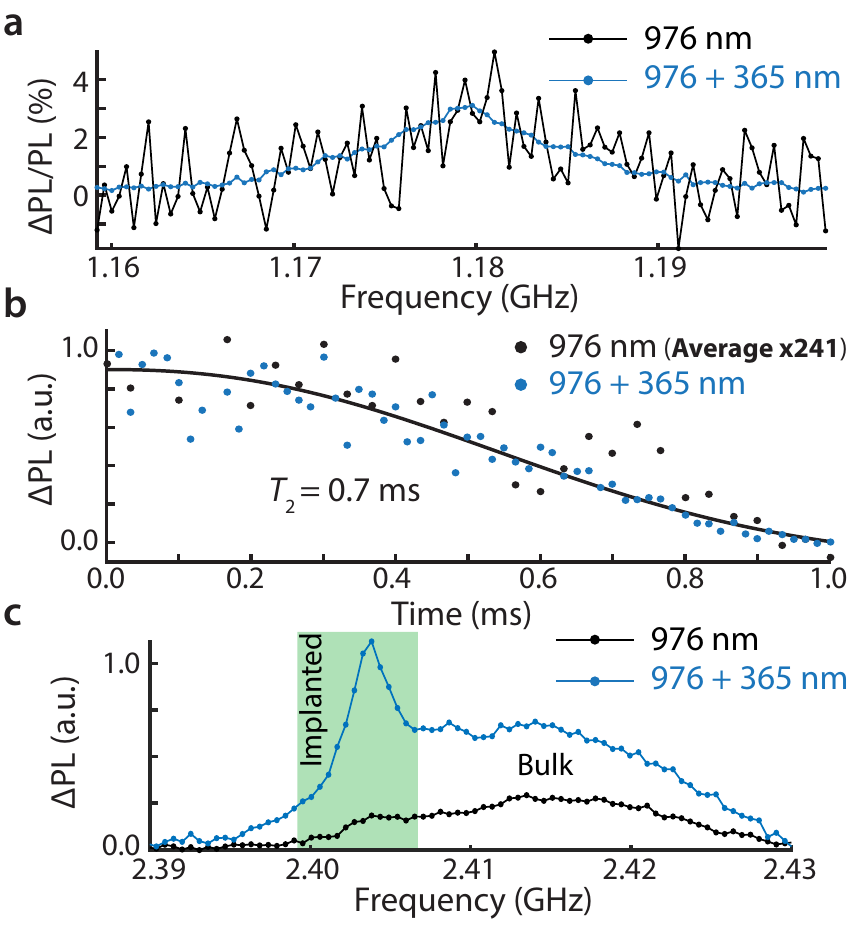}}
	\caption{\textbf{Charge conversion effect on the spin properties.} ODMR signals are given as relative photoluminescence intensities ($\Delta$PL) under microwave excitation.
		{\bf (a)} CW-ODMR spectrum at 50~G and measured through a monochromator at the 1130.6~nm PL2 zero-phonon line to ensure no other contribution in the optical signal. The intensity is given as the ratio (i.e. contrast) between the ODMR and PL intensity, which remains constant with and without 365~nm illumination, indicating unchanged spin polarization and readout mechanisms. 
		{\bf (b)} Hahn echo decay experiment for PL2 at $\approx 400$~G, measured with pulsed-ODMR at 6~K. The 365~nm excitation is continuous throughout the sequence, resulting in a signal increase while the coherence time is unaffected. Decay with 976~nm excitation only was averaged 241 times more than the decay with also 365~nm illumination. Line (in black) is a stretched exponential fit (stretch factor $\approx$~2) to the data.
		{\bf (c)} CW-ODMR (PL2 at $\approx 400$~G) of a 4H-SiC sample with a 500~nm carbon-implant layer below the surface. The divacancies created at the layer are barely visible before 365 nm excitation. The implanted layer peak is also shifted from the bulk due to a magnetic field gradient across the sample. 
	} 
	\label{fig:SpinApp}
\end{figure}

Above-bandgap excitation can be used to efficiently convert VV toward its neutral charge state, and more importantly drastically increases the PL intensity. For practical applications however, we verify this has no effect on the spin properties of \VVO. In \Fig{fig:SpinApp}(a), we first measure the ODMR contrast of PL2, i.e. the ratio of ODMR over PL intensity, which provides a direct measure of how the spin states may be affected during illumination. For these experiments, the 405~nm laser is replaced by a 365~nm (also called "UV") light-emitting diode which is more efficient at charge conversion since it is above bandgap in energy. No difference in contrast is observed with or without 365~nm, and the charge conversion therefore does not significantly affect the spin state nor the readout mechanism. However, the signal-to-noise ratio improves by $\sim$70 times with illumination due to increased \VVO\ charge population. More details are given in the Supplementary Figure 6 and Supplementary Note 2 regarding the ODMR experiments presented here.

A second crucial property is the electron spin coherence of the defect. From \Fig{fig:PLspectrum}(c), the charge stability at cryogenic temperature (6~K) is shown to be much longer than any coherence timescale~\cite{Seo2016}, however we check that even with constant 365~nm illumination ($\sim$0.2~mW) and corresponding electron-hole pair generation, the coherence time remains unaffected. At 400~G, the ensemble electron spin coherence \ttwo\ is measured to be 0.7~ms and remains completely unaffected by either light or free carriers (\Fig{fig:SpinApp}(b)), while much longer averaging ($> \times 200$) was required to obtain similar signal-to-noise ratios without 365~nm illumination. This is not an obvious result as scattering or exchange interaction with free carriers can easily reduce \tone\ or \ttwo\ of defect spins~\cite{Tyryshkin2012}.

Until now, all measurements were realized on as-grown commercial wafers with naturally occurring impurity concentrations. However, carbon ion implantation or electron irradiation~\cite{Falk2013} is often used to increase the PL intensity and to improve spatial resolution. The type of defects created by lattice damage during these processes cannot be well controlled however, though partially manageable using annealing, and the local Fermi level may shift away from obtaining a desired charge state. We test this with a 500~nm thick layer of implanted divacancies (see Methods section). When measuring the PL2 ODMR spectrum of this sample, as shown in \Fig{fig:SpinApp}(c), we obtain a broad ``bulk'' signal observable across the entire sample depth using simply 976~nm excitation. When 365~nm is turned on (with constant absorption over the sample depth), the bulk intensity increases as expected, but more importantly a narrower and more intense peak appears. The latter is assigned to the implanted layer which, being confined in depth, is less sensitive to inhomogeneity in the static magnetic field. Rabi experiments at the peak layer frequency yielded as expected an increased contrast (Supplementary Figure 7), demonstrating that the UV charge stabilization can be critical in such samples. 

\subsection{Charge state conversion}
\begin{figure*}[t]
	\centerline{\includegraphics[width=\textwidth]{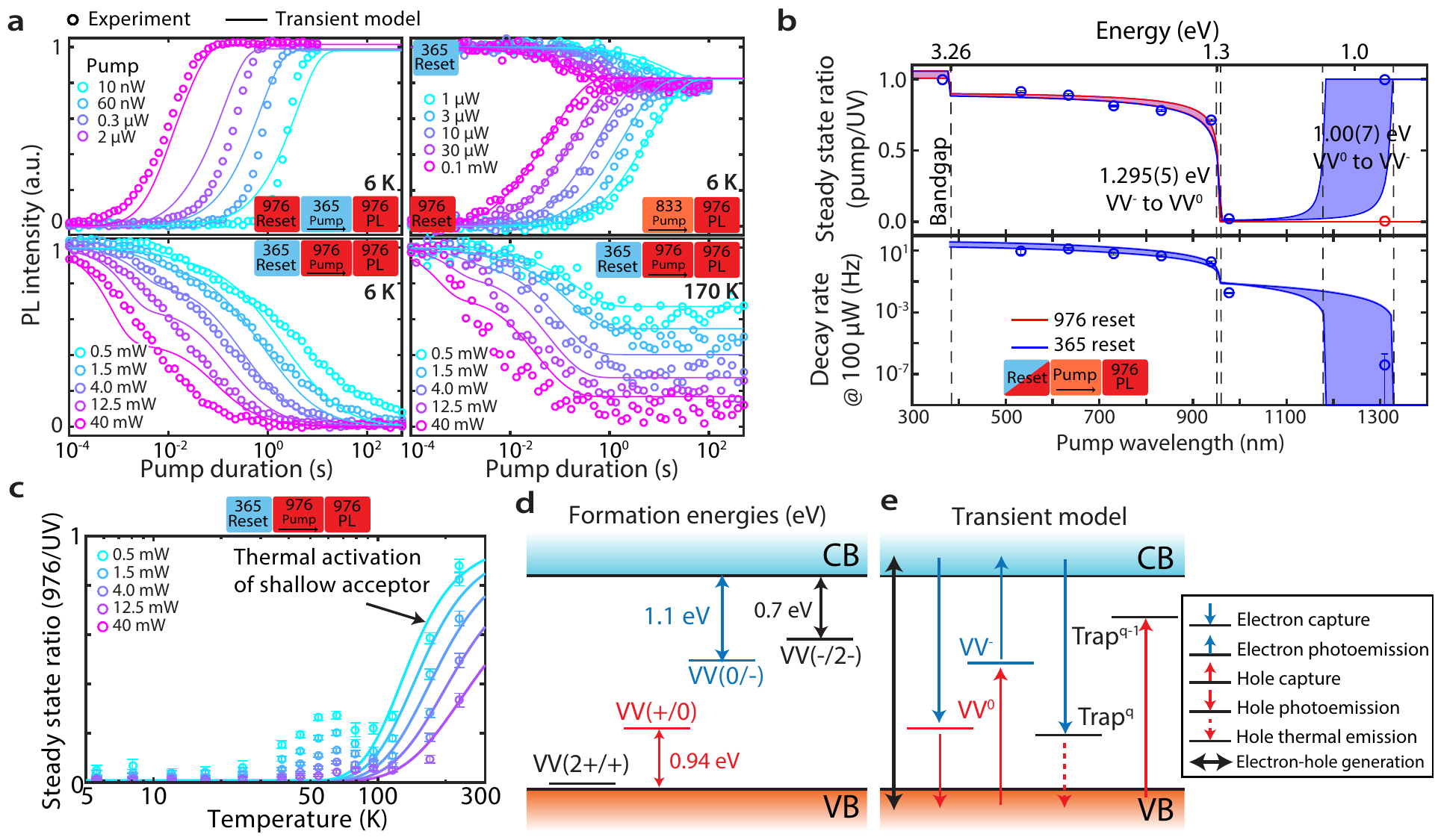}}
	\caption{\textbf{Photo-dynamics and modeling in neutral divacancies in 4H-SiC.} The charge dynamics are probed using two and three color experiments, following a reset-pump-measure scheme.
		{\bf (a)} Typical decay curves obtained under various reset/pump wavelength and temperatures. The fit (line) is obtained from the model given in (e).
		{\bf (b)} Top figure: ratio between pump and 365~nm steady state PL intensities. Bottom figure: decay rates (normalized to 100~$\mu$W at every pump wavelengths) obtained by fitting the decays in (a) with a stretch-exponential function. In blue, the sequence starts after 365~nm pumping, while in red, after 976~nm. The lines are given by the model in (e), with the area corresponding to 95\% confidence intervals. For 1310~nm, no significant decay was observed over 100~s, hence the steady-state values are given without errorbars.
		{\bf (c)} Temperature dependence of the steady state after 976~nm pumping (365~nm reset). Lines are given by the model in (e), corresponding to thermal activation of a shallow acceptor. The origin of the intermediate region between 30~K and 100~K is unknown. Above 210~K, PL5 and PL6 signals become dominant, making the measurement unreliable as they are UV-insensitive.
		{\bf (d)} Formation energies of the divacancy in 4H-SiC, taken from \cite{Gordon2015}.
		{\bf (e)} Model used for simulating all transients in (a), (b) and (c), including the \VVO\ and \VVm\ levels of the divacancy, as well as an unknown trap with two charge states. Processes included in the model are given in the legend.
	} 
	\label{fig:Transients}
\end{figure*}
We now consider in more depth the charge mechanisms within 4H-SiC, in particular the effect of 976~nm illumination which appears to convert VV toward a dark charge state (\VVm\ or \VVp). Since 976~nm is used both for PL excitation and causes charge conversion, it is necessary to separate the two contributions from the PL intensity, which can be achieved by looking at the conversion dynamics under pulsed light. This is realized using a three-pulse scheme: reset with either 976~nm or 365~nm, pump with a wavelength ranging from 365~nm to 1310~nm using various laser diodes, and measurement with 976~nm. Typical decay curves as a function of pumping duration are shown in \Fig{fig:Transients}(a) with different initial reset lasers, pump wavelength, pump power and temperatures.

In \Fig{fig:Transients}(b), fitted steady states and decay rates (see Methods section regarding the fitting) are plotted as a function of pump excitation wavelength. Steady-state intensities are all normalized by the steady-state PL intensity after UV pumping. A clear transition is observed between 940~nm and 976~nm, at about 1.3~eV, for both steady-state values and decay rates, with shorter wavelengths being increasingly more efficient at charge conversion toward \VVO. In addition, we measure a single wavelength at 1310~nm that tentatively suggests a second transition (between 976~nm and 1310~nm), where the \VVO\ charge state becomes insensitive to excitation (no observed decay). The wavelength transitions can be related to photoionization energies and, though the Franck-Condon shift is unknown here, to formation energies obtained from DFT calculations~\cite{Gordon2015} and reproduced in \Fig{fig:Transients}(d). The divacancy defect in 4H-SiC has four stable charge states: $+$, $0$, $-$ and $2-$. The $(+/0)$ and $(0/-)$ transition levels are calculated to be, respectively, $\sim$~\Ev\ + 0.94~eV and $\sim$~\Ec\ - 1.1~eV, with \Ev\ and \Ec\ the valence and conduction band energies. Considering typical uncertainty in DFT calculations of 0.1~eV as well as Franck-Condon shift in the order of 0-0.3~eV (for \Vc\ for example \cite{Son2012}), this matches fairly well with our measured values and the fact that the (0/-)/940~nm transition is at higher energy than (0/+)/976~nm. We can therefore accredit our results to charge conversion between the \VVO\ and \VVm\ states.

We attempt to model the observed dynamics using the rate-equation model shown in \Fig{fig:Transients}(e), based on charge transfer between the divacancy and a trap of unknown origin. More details are given in both the Methods section and Supplementary Note 1. Simulated decays and their corresponding rates and steady states are shown in \Fig{fig:Transients}(a,b). Three experimental characteristics are nicely reproduced by the model here: i) the jump in charge conversion efficiency for above bandgap illumination, ii) the \VVm\ to \VVO\ transition fitted to be $E_{-0} = 1.295(5)$~eV and iii) the \VVO\ to \VVm\ transition roughly estimated at $E_{0-} = 1.00(7)$~eV (1150 to 1350~nm). For \VVO\ charge stability, an important parameter is the ratio of cross-sections between the electron (\VVm\ $\rightarrow$ \VVO) and hole (\VVO\ $\rightarrow$ \VVm) photoionization processes, which roughly follows the relation $K_{\sigma}\left(\frac{E-E_{-0}}{E-E_{0-}}\right)^{3/2}$ with $K_{\sigma} = 26 (8)$ (valid at or above $E_{-0}$). \VVO\ is therefore the stable charge state for any illumination above $\approx 1.3$~eV at cryogenic temperatures. 

Finally, a temperature dependence of 976~nm pumping (365~nm reset) is taken between 5.5~K and 210~K, with corresponding steady states value shown in \Fig{fig:Transients}(c). Above 100~K, the effect of 976~nm pumping compared to 365~nm pumping is drastically reduced. The simulation is able to reproduce this feature owing to hole thermal emission from the trap, with a tentative activation energy between 0.05 and 0.15~eV depending on the fitting conditions, such as which temperatures to include in the data set.

In summary, we identify a sharp transition of the VV charge dynamics at around 1.3~eV (960~nm), corresponding to the ionization of \VVm to \VVO. Below 1.3~eV in energy, hole photoemission drives the charge state toward \VVm, while above, VV is preferentially in \VVO and remains stable for many hours after illumination. In addition, UV light above bandgap strongly drives the system toward \VVO.

\subsection{Charge transfer between major defects}
The experiments described previously made use of PL as a direct measurement of the divacancy neutral charge state, combined with photo-excitation to probe relevant energy levels as well as trapping or recombination dynamics. However, understanding all the major defects in 4H-SiC, not just the divacancy, is required to obtain a comprehensive picture of the sample behavior under illumination. While PL of \Vsi\ can be measured, other important spin impurities such as \Vc\ or N are not photo-active, with no optical excited states in the bandgap. We thus turn toward electron spin resonance (ESR) to provide information on all the spin species. 

\begin{figure}[t]
	\centerline{\includegraphics[width=\figwidth]{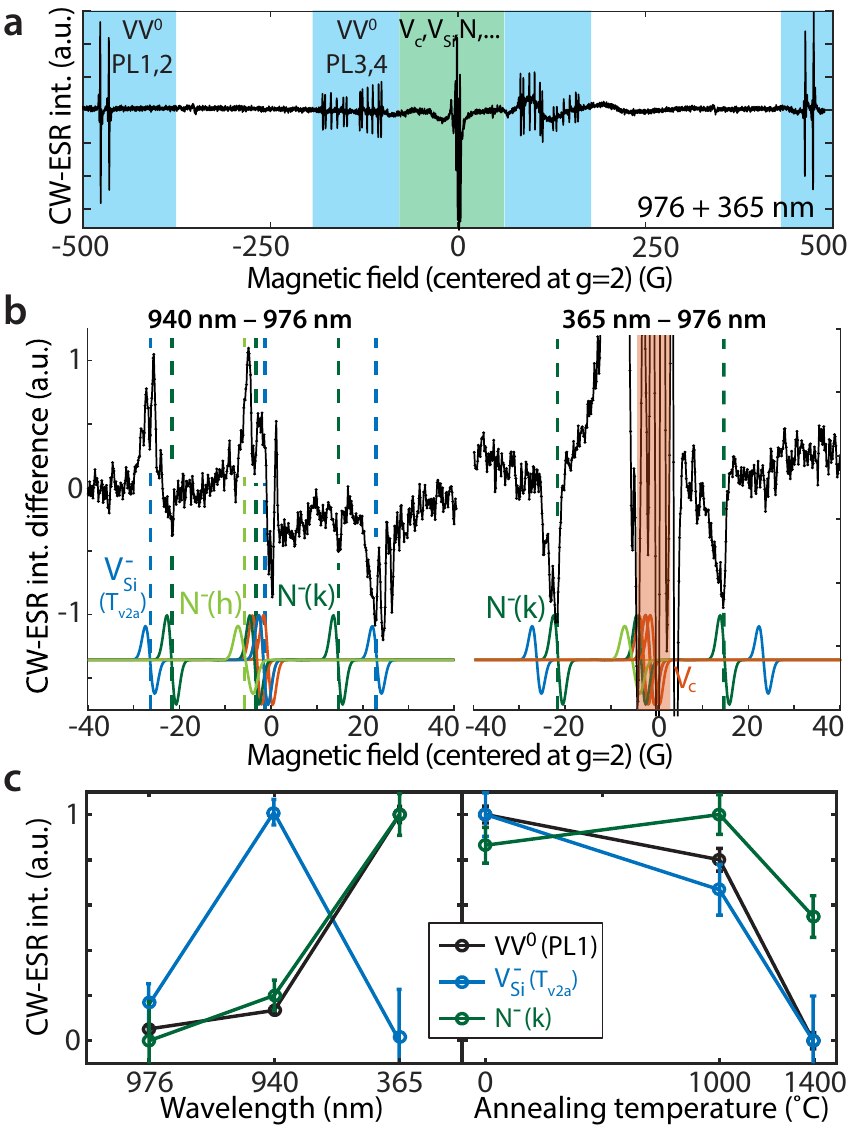}}
	\caption{\textbf{ESR at 15~K in semi-insulating 4H-SiC under illumination.} 
		{\bf (a)} CW-ESR spectrum measured at 9.7~GHz, and centered around g=2 ($\approx$~3470~G, aligned to the c-axis). VV PL1-4 are highlighted in blue, while defects such as N, \Vc\ or \Vsi\ are close to $g=2$ and highlighted in green.
		{\bf (b)} Differential CW-ESR spectrum between either 940~nm and 976~nm excitation (left), or between 365~nm and 976~nm (right). Gaussian derivative lineshapes are simulated in color for known defects in 4H-SiC~\cite{Isoya2008}. Their amplitudes only take into account transition probabilities, and not spin polarization or microwave saturation.
		{\bf (c)} Normalized (per defect) CW-ESR intensity under 976~nm, 940~nm and 365~nm (left) and for different annealing condition of the sample (right). For \VVO, 365~nm is combined with 976~nm for spin polarization (and obtain enough signal). For the annealing dependence, the intensity was fitted under the best illumination condition for each defect, that is the maximum signal in the left panel. Annealed samples were only used in this panel ((c), right).
	} 
	\label{fig:ESR}
\end{figure}

A CW-ESR spectrum at X-band is shown in \Fig{fig:ESR}(a) with resonance peaks from PL1 to PL4 VV defect types, as well as a cluster of signals near the g-factor $g=2$, known to be from \Vsi,\Vc\ and/or N~\cite{Son2006,Isoya2008}. In order to properly resolve some of these peaks, we subtract the ESR spectrum measured with 976~nm illumination from the spectrum obtained with either 940~nm or 365~nm illumination. The differential spectra, shown in \Fig{fig:ESR}(b), then correspond to possible charge transfer with VV which is extremely sensitive to these wavelengths (further considerations are discussed in the Methods section). The ESR peak intensities are given for the main identified defects in \Fig{fig:ESR}(c) (left). With 940~nm, two sets of resonances can be clearly assigned, the strongest due to \Vsim\ ($T_{V2a}$ or $V2$ center) and the weaker from \Nm\ (k site). At this wavelength, \VVm\ undergoes photoionization to become \VVO, emitting an electron to the conduction band which is likely captured by \VsiO, and resulting in an increase in \Vsim. With 365~nm, large changes can be seen around $g=2$, possibly from free electrons and \Vc, though the peaks are too clustered to be resolved. On the side of $g=2$, \Nm\ appears much stronger while the \Vsim\ peaks completely disappeared. While this may be due to carrier-induced spin relaxation, such behavior is also well explained by charge dynamics: N, initially in its neutral charge state due to either photoionization or thermal emission (shallow donor) before cooling down the sample, captures most of the generated electrons to give a high \Nm\ signal. The holes now in majority, are captured by the various deep defects, with \VVm\ being converted toward \VVO\ (high signal), \Vsim\ toward \VsiO\ (low signal), and possibly \VcO\ toward \Vcp\ (high signal).

ESR measurements are often used to characterize the optimal annealing temperature for sample preparation. Changes in ESR intensity after sample annealing normally indicates variations in defect concentration, but can also be confused with a shift in the local Fermi level. After charge conversion, this second explanation is much less plausible. In \Fig{fig:ESR}(c), the ESR intensity under best illumination condition (highest signal for each defect) was tracked for N, \Vsi, and VV for different annealing temperatures. Between 1000~\degree C and 1400~\degree C, the ESR signal of \Vsi\ and VV significantly drops, which can be related to the defects becoming mobile followed by creation of multi-vacancies such as \Vc-\Vsi-\Vc~\cite{Gerstmann2003, Zolnai2004, Schmid2006, Carlos2006}.

\begin{figure}[t]
	\centerline{\includegraphics[width=\figwidth]{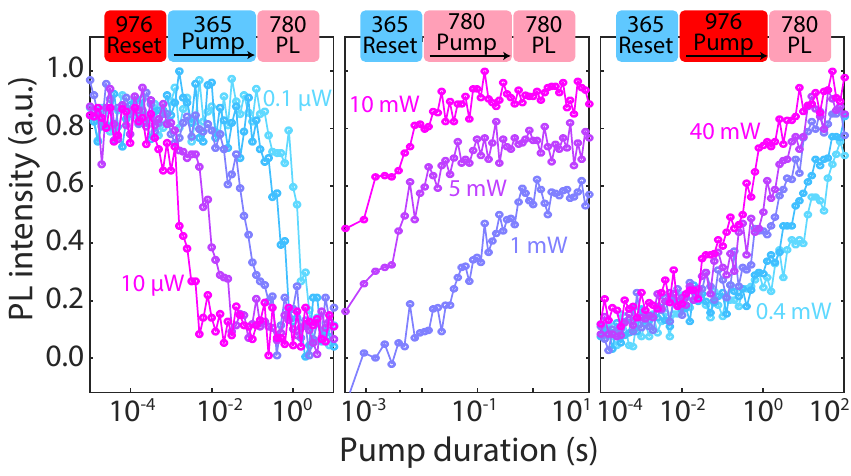}}
	\caption{\textbf{Photo-dynamics of \Vsim\ at 6~K.} 
		Reset-pump-measure scheme similar to \Fig{fig:Transients}, but with 780~nm to excite PL in \Vsim\ instead of 976~nm for \VVO. 365~nm reduces the PL intensity, likely from charge conversion to \VsiO. Charge conversion was measured to be persistent without light on the experiment timescales.
	} 
	\label{fig:Vsi}
\end{figure}

The ESR experiments indicate a strong relationship between VV and \Vsi, as they are both affected by the 940-976~nm transition and by similar annealing temperatures. With \Vsim\ being also a photo-active qubit of interest, we directly measure its PL by exciting the sample with 780~nm (see Supplementary Figure 1)~\cite{Baranov2011, Embley2017}. Three-pulse experiments for \Vsim\ are shown in \Fig{fig:Vsi} with pumping using 365~nm, 976~nm, as well as 780~nm as a replacement for 940~nm which was impractical here. 365~nm pumping drastically decreases the \Vsim\ intensity while both 976~nm and 780~nm convert back the charge state to \Vsim, with 976~nm illumination being less effective. These observations are consistent with the ESR experiments. The charge conversion toward \Vsim\ is ascribed to \VsiO\ hole photo-emission, with the difference in conversion efficiency resulting from VV photo-emission. Indeed, VV emits holes under 976~nm and dominantly electrons under 780~nm, which has opposite charge effects for \Vsi. For example, having more electrons with 780~nm excitation pushes \Vsi\ faster toward its higher charge state \Vsim.

Looking at formation energies for \Vsi\ (\cite{Hornos2011,Gordon2015}, reproduced in Supplementary Figure 5), the \Vsi$(+/0)$ transition is very shallow (\Ev\ + 0-50~meV) and can be photo-excited at any wavelength while \Vsi$(0/-)$ is near mid-gap (\Ev\ + 1.3-1.5~eV). The temperature dependence and transient modeling in \Fig{fig:Transients} matches with the presence of \VsiO\ as a shallow acceptor that can trap and re-emit holes. The activation energies from theory and simulation are similar (0.05-0.15~eV), though with very large uncertainties in both cases. For completeness, the \Vsi$(-/2-)$ transition with a formation energy of \Ec\ - 0.6-0.8~eV is likely also excited at any wavelength below 976~nm, and therefore \Vsi\ is more likely to be trapped in \Vsim\ than in the $2-$ charge state. 780~nm illumination is therefore suitable for both \Vsim\ PL excitation and \Vsim\ charge stabilization.

\begin{figure}[t]
	\centerline{\includegraphics[width=\figwidth]{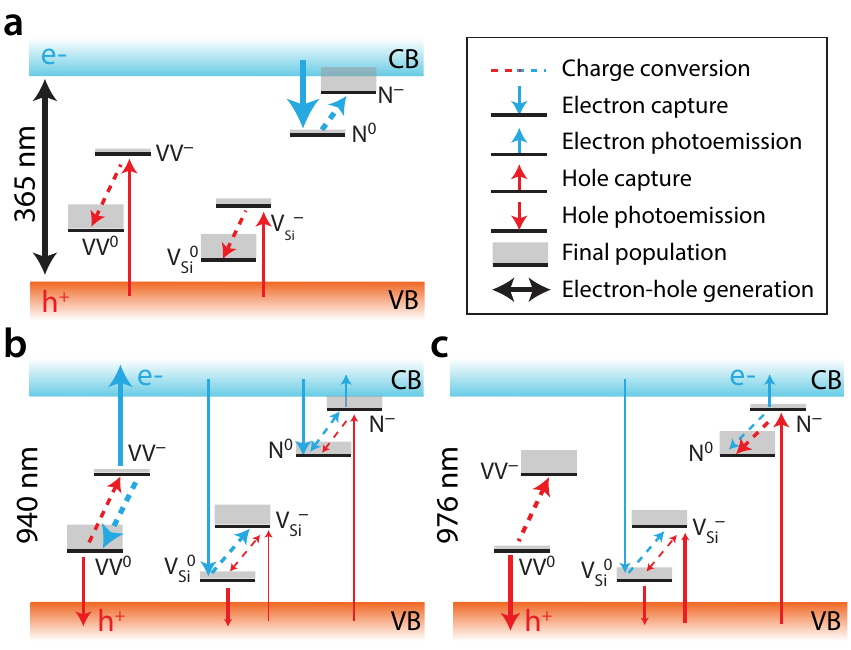}}
	\caption{\textbf{Summary of charge transfer in semi-insulating 4H-SiC under various illumination conditions.} Strong transitions are shown by thicker arrows, and the steady-state population after illumination is approximatively represented by the grey area over each state.
		{\bf (a)} Above-bandgap excitation and electron-hole generation.
		{\bf (b)} Excitation above the \VVm\ photoionization transition ($\sim$1.3~eV).
		{\bf (c)} Excitation below the \VVm\ photoionization transition, but above the \VVO\ hole emission.
	} 
	\label{fig:Summary}
\end{figure}

The full VV/\Vsi/N charge conversion picture under illumination is finally summarized in \Fig{fig:Summary} for the three critical wavelengths explored in this work: 976~nm, 940~nm and 365~nm. \Vc\ was not taken into account due to lack of measurements, but it is likely trapped to \Vcp\ or \VcO\ where it is too deep to be photo-excited (\cite{Hornos2011,Gordon2015}, reproduced in Supplementary Figure 5).

\subsection{Summary of charge dynamics}
Our overall summary of the charge conversion is as follows: i) Under 365~nm excitation and electron-hole generation, N dominantly traps electrons toward \Nm, while VV and \Vsi\ capture the remaining holes to become \VVO\ and \VsiO. ii) Below but close to 940~nm in wavelength, VV chiefly emits electrons and ends up in \VVO, while \Vsi\ both captures those electrons and emits holes to become \Vsim. N will be in an intermediate charge state as it absorbs both electron and holes, as well as being slighly photoionized. iii) At wavelengths higher than 976~nm, VV is converted to \VVm\ by hole emission; N and \Vsi\ then both capture those holes and are photoionized, resulting in \NO\ and a slow conversion toward \Vsim. 

The charge conversion and transfer mechanisms presented throughout this work should remain valid in most semi-insulating materials, where defects are in comparable concentrations. For n- or p-doped materials, impurities can of course still be photoionized, however electron-hole generation with above-bandgap light will likely set the local Fermi level to a different equilibrium than what is seen here.

\subsection{Toward applications: charge patterning}
To complete this study, we turn toward applications using our ability to control the VV charge state. In recent experiments in diamond~\cite{Dhomkar2016}, optical conversion between the NV$^{-}$ and NV$^{0}$ states was used to demonstrate the possibility of information storage by 3D patterning of the charge state. Because data can be both encoded in 3D as well as a gradient of charge conversion, high storage densities can theoretically be achieved. We present a similar demonstration of charge patterning in 4H-SiC, and though our experiments are realized at 6~K, offer the potential for storage across entire wafers compared to diamond. The VV charge conversion works relatively well up to 150-200~K, and may possibly be extended to room temperature with the adequate choice of material (dominant dopant or impurity concentration). 

\begin{figure}[t]
	\centerline{\includegraphics[width=\figwidth]{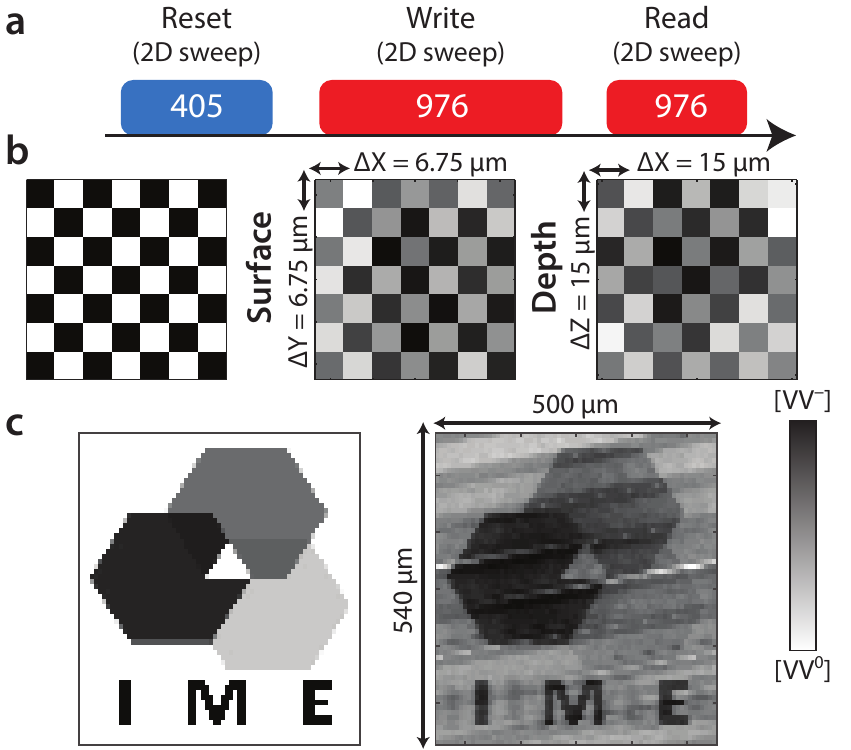}}
	\caption{\textbf{Spatial and amplitude control of the divacancy charge conversion.}
		{\bf (a)} Imaging sequence for (b) and (c), with three sequential 2D sweeps: 1) Reset to high VV$^0$ concentration using 405~nm. 2) Write using 976~nm with varying duration for charge conversion back to a desired lower VV$^0$ concentration. 3) Read with a fast 976~nm pulse.
		{\bf (b)} Pixel test of spatial control with, from left to right, the original pattern, the measured pattern in the X-Y plane (parallel to the sample surface) and the measured pattern in the X-Z plane (orthogonal to the sample surface). For X-Z, the intensity is normalized to the PL collection efficiency across the sample depth.
		{\bf (c)} Amplitude control of the charge conversion using a gray scale image (left: original image, right: experiment).
	} 
	\label{fig:Imaging}
\end{figure}

The patterning scheme is presented in \Fig{fig:Imaging}(a) with: a UV (405~nm) pulse to initialize the sample toward \VVO, a write pulse with 976~nm to selectively obtain \VVm, and finally a short read pulse using 976~nm. The measurement pulse here weakly destroys the information due to undesired charge conversion, which is the main limitation to this technique. In \Fig{fig:Imaging}(b), we test the spatial resolution of our setup by patterning a pixelized checkerboard design (left), first parallel to the sample plane (middle) and then in depth, orthogonal to the sample plane (right). Finally, for each pixel, we allow control over the amount of charge conversion, increasing the density of information that can be locally stored. This is demonstrated in \Fig{fig:Imaging}(c) by patterning a 500~$\mu$m~$\times$~540~$\mu$m grey scale image parallel to the sample surface.

\section{Discussions}
In this work we have systematically investigated the charge properties of divacancies in semi-insulating 4H-SiC, as well as other relevant defects such as \Vsi\ and N. Through optical excitation with wavelengths spanning from 365 to 1310~nm, VV was found to be stable in either its neutral or negative charge configuration. The photoionization energies of both \VVO\ and \VVm\ are fitted to be around 1.3~eV and 1~eV; in particular, the commonly used 976~nm excitation for PL measurements is found to be detrimental as it converts VV toward \VVm. Above-bandgap excitation efficiently reshuffles the charge states of all defects, with VV becoming bright (\VVO) and \Vsi\ becoming dark (\VsiO). Overall, taking into account all impurities was necessary to obtain a complete picture of charge effects in these samples; such considerations are crucial for tuning wafer growth techniques, samples with implanted layers, surface impurities or for devices with complex electric potentials. Finally, we confirmed that these optical charge conversions drastically improve the PL intensity and do not impact in any way the spin properties (ODMR, coherence). Combined with recent studies \cite{Christle2017,Fuchs2015} characterizing the spin and optical properties of VV or \Vsi\ in 4H and 3C-SiC, this work on charge conversion/stabilization helps to complete the suite of techniques and technologies realized in NV centers in diamond for use in SiC, while allowing for novel applications such as optically controlling the charge of spins in electronic devices realized in SiC.

\section{Methods}
\subsection{Samples}
All measurements were performed on commercially available high-purity semi-insulating 4H-SiC diced wafers purchased from Cree~\cite{Jenny2004}, and using a scanning ODMR microscopy setup. Similar wafers have been used in other studies, with measured defect concentrations of N, VV, \Vsi\ all in the order of $10^{14}-10^{16}$~cm$^{-3}$~\cite{Jenny2004,Son2007,Chandrashekhar2012}. For the implanted sample, a high energy carbon implant ($[^{12}\rm{C}] = 10^{13}\ \rm{cm}^{-2} , 190$~keV, 900\degree C anneal for 40~mn) was used, resulting in a calculated (SRIM software) 500~nm thick layer of divacancies. For ODMR, the samples are fixed to a printed circuit board patterned with a coplanar waveguide for magnetic resonance, and mounted in a closed-cycle cryostat cooled down to 5-6~K (unless otherwise mentioned). PL, ODMR and ESR experiments were all realized on ensembles of defects.

\subsection{PL and ODMR setups}
For \VVO, the sample is excited with a 976~nm diode laser (40~mW at sample, focused with a 50X IR objective) and PL is measured with an InGaAs detector (1000-1300~nm after filtering). For \Vsim, the sample is excited with a 780~nm diode laser (10~mW at sample) and PL is measured with a Si detector (850-950~nm after filtering, allowing simultaneous \VVO\ PL recording). Simplified schematics for the PL/ODMR setups are given in Supplemental Figure 1. All given optical powers were measured at the sample. Excitation spectra are recorded by inserting a monochromator immediately after a 100~W Xe white light source in the optical setup. Emission spectra or measurements at selective zero-phonon lines (ZPL) are recorded by inserting a monochromator before the detector. For the wavelength dependence, a set of laser diodes were successively collimated into a 300~$\mu$m multi-mode fiber and re-emitted into free space so as to ensure a constant spot position on the sample. It should also be noted there is no significant PL contribution from 405~nm or 365~nm illumination alone.
 
\subsection{Transients and modeling}
The three-pulse scheme used for the photo-dynamics requires careful choice of the 976~nm measurement pulse duration (0.1~ms) as it is necessary for exciting PL but can also change the charge state of VV. A long pulse would effectively smooth the decays and prevent good fitting at short times. The experimental decay rates and steady states are obtained from fitting with a stretched exponential function, with separate fitting parameters for each power, wavelength and temperature dependence. The actual decay curves are shown in Supplementary Figure 3, and all simulated lines in \Fig{fig:Transients} are from the rate-equation model. All details on this model are given in Supplementary Figure 4 and Supplementary Note 1, regarding e.g. simulation of the wavelength dependence (Grimmeiss model for deep trap~\cite{Grimmeiss1975}) and the exact rate equations. 

In total, 11 parameters are used for a simultaneous fit over a set of 70 decays curves, with the simulation results shown by the lines in \Fig{fig:Transients}(a), (b) and (c). Looking at the decays in \Fig{fig:Transients}(a), the fits are in excellent agreement in certain ranges (833~nm pumping) but do not account for all the charge dynamics as seen in the left figures. For 365~nm pumping, electron-hole pair generation dominates over all photoionization processes, and the free carrier concentration is determined by the recombination with all involved traps, not just VV. Hence for such a simple model, large discrepancies are expected. In addition, the simulation strictly considers a single defect while measuring an ensemble can easily smooth features in the decays, e.g. due to local variations in strain, charge, light intensity, etc.

\subsection{Electron spin resonance}
ESR experiments were realized on a X-band (dielectric resonator, 5~mm internal diameter) ELEXSYS E580 Bruker spectrometer at 15~K. In the differential experiments presented in \Fig{fig:ESR}, one important issue is the simultaneous effect of illumination on both spin (polarization, relaxation) and charge properties, and hence on the ESR intensity. The results presented may convolute both aspects, unlike the PL experiments which are clearly related to the charge state. Turning the lasers on and off would avoid this concern, however the signal was then simply too weak to obtain any information on N, \Vc\ or \Vsi. The 940 and 976~nm excitation lasers are sufficiently close in energy to limit most effects but those related to the sharp photoionization transition in VV. In addition, these wavelengths are both in the VV absorption sideband, but close enough to see no appreciable differences in spin polarization due to inter-system crossing mechanisms. Similarly, they are also above the longest ZPL wavelength of \Vsim\ (917~nm~\cite{Baranov2011}), preventing any spin-polarization.

\section{Acknowledgments}
We thank Adam Gali, Hosung Seo, Alexandre Bourassa and David Christle for discussions. G.W. acknowledges support from the University of Chicago/Advanced Institute for Materials Research (AIMR) Joint Research Center. A.L.Y, F.J H., and D.D.A. were supported by the US Department of Energy, Office of Science, Basic Energy Sciences, Materials Science and Engineering Division at Argonne National Laboratory. J.N. and O.G.P. were supported by the U.S. Department of Energy, Office of Science, Office of Basic Energy Sciences, Division of Chemical Sciences, Biosciences and Geosciences under Contract DE-AC02-06CH11357 at Argonne National Laboratory. C.P.A. was supported by the Department of Defense (DoD) through the National Defense Science and Engineering Graduate Fellowship (NDSEG) Program.

\section{Author contributions}
G.W., C.P.A. and A.L.Y. performed the optical experiments. G.W., J.N., O.G.P. and F.J.H. performed the electron spin resonance experiments. S.J.W. and C.P.A. processed the annealed samples and A.L.Y. designed the implanted sample. All the authors contributed to analysis of the data, discussions and the production of the manuscript.

\bibliography{./library}

\end{document}

%% file: Defs.tex
\usepackage{graphicx}
\graphicspath{{./Figs/}}
\usepackage{color}
\usepackage{bm}
\usepackage{dsfont}
\usepackage{epstopdf}
\usepackage{hyperref}

\newcommand {\figwidth} {3.4in}

\newcommand {\Fig}[1] {Figure~\ref{#1}}

\newcommand {\degree} {$^{\circ}$}

\newcommand{\ttwo}{$T_2$}
\newcommand{\tone}{$T_1$}

\newcommand{\Ec}{$E_{\rm c}$}
\newcommand{\Ev}{$E_{\rm v}$}

\newcommand{\VVO}{VV$^0$}
\newcommand{\VVp}{VV$^+$}
\newcommand{\VVm}{VV$^-$}

\newcommand{\Vsi}{V$_{\rm Si}$}
\newcommand{\VsiO}{V$_{\rm Si}^0$}
\newcommand{\Vsim}{V$_{\rm Si}^-$}

\newcommand{\Vc}{V$_{\rm C}$}
\newcommand{\VcO}{V$_{\rm C}^0$}
\newcommand{\Vcp}{V$_{\rm C}^+$}
\newcommand{\Nm}{N$^-$}
\newcommand{\NO}{N$^0$}